\documentclass[conference]{IEEEtran}
\IEEEoverridecommandlockouts

\usepackage{cite}
\usepackage{amsmath,amssymb,amsfonts}
\usepackage{algorithmic}
\usepackage{graphicx}
\usepackage{textcomp}
\usepackage{xcolor}
\usepackage{booktabs}

\newcommand{\todo}[1]{}
\renewcommand{\todo}[1]{{\color{red} TODO: {#1}}}

\newcommand{\rmm}[1]{}
\renewcommand{\rmm}[1]{{\color{gray} RM: {#1}}}
\usepackage{comment}
\usepackage{tikz}
\usepackage{multirow}
\usepackage{xcolor,colortbl}
\usepackage{makecell}
\usepackage{diagbox}
\usepackage{nicematrix}
\usepackage{url}

\definecolor{C0}{RGB}{31,119,180}  
\definecolor{C1}{RGB}{255,127,14}  
\definecolor{C2}{RGB}{44,160,44}   

\usetikzlibrary{positioning, arrows.meta, calc}
\usetikzlibrary{bayesnet}
\usetikzlibrary{arrows}

\def\BibTeX{{\rm B\kern-.05em{\sc i\kern-.025em b}\kern-.08em
    T\kern-.1667em\lower.7ex\hbox{E}\kern-.125emX}}
\begin{document}

\title{DeCRED: Decoder-Centric Regularization for Encoder-Decoder Based Speech Recognition
}


\author{
\IEEEauthorblockN{
Alexander Polok, Santosh Kesiraju, Karel Beneš, Bolaji Yusuf, 
Lukáš Burget, Jan Černocký}
\IEEEauthorblockA{Brno University of Technology, Brno, Czech Republic}
}

\maketitle

\begin{abstract}
This paper presents a simple yet effective regularization for the internal language model induced by the decoder in encoder-decoder ASR models, thereby improving robustness and generalization in both in- and out-of-domain settings.
The proposed method, Decoder-Centric Regularization in Encoder-Decoder (DeCRED), adds auxiliary classifiers to the decoder, enabling next token prediction via intermediate logits.
Empirically, DeCRED reduces the mean internal LM BPE perplexity by 36.6\,\% relative to 11 test sets.
Furthermore, this translates into actual WER improvements over the baseline in 5 of 7 in-domain and 3 of 4 out-of-domain test sets, reducing macro WER from 6.4\,\% to 6.3\,\% and 18.2\,\% to 16.2\,\%, respectively.
On TEDLIUM3, DeCRED achieves 7.0\,\% WER, surpassing the baseline and encoder-centric InterCTC regularization by 0.6\,\% and 0.5\,\%, respectively.
Finally, we compare DeCRED with OWSM v3.1 and Whisper-medium, showing competitive WERs despite training on much less data with fewer parameters.
\end{abstract}


\section{Introduction}

One of the key challenges in automatic speech recognition (ASR) is to enhance the ability of models to generalize to new or unseen domains. Regularization techniques have long been central to improving robustness. Methods such as SpecAug~\cite{park_specaugment_2019}, label smoothing~\cite{PereyraTCKH17,kim18f_interspeech}, multi-task training~\cite{hori_joint_2017}, and architecture-specific regularization~\cite{lee_intermediate_2021} are now standard practices in ASR pipelines and are integrated into open-source toolkits~\cite{watanabe_espnet_2018,speechbrain}. Another prominent strategy is large-scale multi-domain training~\cite{narayanan_toward_2018}, as demonstrated in works like SpeechStew~\cite{chan_speechstew_2021}, Canary~\cite{puvvada24_interspeech}, Whisper~\cite{radford_robust_2023}, OWSM~\cite{peng_reproducing_2023}, and OWLS~\cite{chen2025owlsscalinglawsmultilingual} which aggregate datasets from diverse domains (e.g., conversational, broadcast, telephone, and read speech) to train models capable of handling broad variability in data.

Although Whisper achieves impressive recognition accuracy, the lack of transparency about its training data has prompted the scientific community to develop open-source equivalents like OWSM~\cite{peng_reproducing_2023}, trained on publicly available datasets.
While scaling up training data has clear benefits, it is computationally expensive and inaccessible to many researchers.
Moreover, it is hard to evaluate the out-of-domain generalization of these models since all the standard datasets were already seen during the training.
Together, these prompt the question:
``Are there any other simpler yet effective techniques to improve robustness?"


In this paper, we propose a straightforward approach to improve out-of-domain generalization by regularizing the auto-regressive Internal Language Model (ILM)~\cite{zeineldeen_investigating_2021,Zhao:2025:ICASSP_ILM_E2E} in encoder-decoder-based ASR architectures.
The use of auxiliary classifier(s) or intermediate regularizers has been explored in ASR, though most prior work applies these techniques to the encoder module.
For example, \cite{lee_intermediate_2021} employs intermediate CTC objectives in the ASR encoder, while \cite{Nozaki_self_cond_2021} extends this idea by feeding intermediate classifier outputs to subsequent layers, conditioning the predictions of the final layer on these intermediate outputs. Similarly, \cite{wang2021selfsupervised} applies such schemes in self-supervised training of speech encoders. More recently, \cite{zhang2022intermediatelayer} regularizes both the encoder and the decoder by passing the intermediate encoder output directly to the decoder layers.
\begin{figure}[!t]
\centering
\begin{tikzpicture}
[
rect/.style={minimum size=0.2cm,minimum height=1.0cm,text width=28mm,
	align=center, rectangle,draw,rounded corners,thick, fill=blue!15},
rect2/.style={rectangle,minimum size=7mm,text width=28mm,   
    align=center,draw,rounded corners,fill=red!20},
emb/.style={minimum size=0.2cm,minimum height=0.5cm,text width=13mm,
    align=center, rectangle,draw,rounded corners,thick, fill=green!20},
shaded/.style={minimum size=0.1cm, minimum height=6mm, text width=0.5mm, align=center,   
	rectangle,draw,thick,fill=lightgray!50},
var/.style={text width=2.4cm,minimum height=8mm,rectangle,align=center},
eqn/.style={text width=8.5cm,minimum height=8mm,rectangle,align=center},
arr/.style={->,>=stealth',semithick},
outer/.style={draw, thick, densely dashed, fill=white!5,
	inner xsep=1ex, inner ysep=0ex, yshift=1ex,	fit=#1},
hidden/.style={circle,scale=0.05,minimum size=1pt,draw},
]
    \tikzstyle{every node}=[font=\footnotesize]

\node (x)  [minimum size=2cm,var]  at (-0.1, -0.2)  {FBANKS $(\mathbf{x}_{1:T})$};
\node (enc) [rect]   at (-0.1, 1.0) {Speech Encoder \\ (E-Branchformer)};
\node (ctc) [emb]    at (-0.1, 2.4) {$\mathrm{CTC}$};
\node (pctc) [var]   at (-0.1, 3.5) {$\mathcal{L}^{\text{CTC}}$};

\node (dec0) [rect2] at (3.5, 0.5) {First Decoder layer};
\node (dec1) [rect2] at (3.5, 1.7) {$(D\text{-}2)^{\textrm{th}}$ Decoder layer};
\node (h13) [hidden] at (3.5, 2.3)   {};
\node (sf1) [emb]    at (5.3, 4.0) {$\mathrm{softmax}$};
\node (pd1) [var]    at (5.3, 5.1) {$\mathcal{L}_{(D-2)}^{\text{Attn.}}$};
\node (dec2) [rect2] at (3.5, 3.0) {$(D)^{\textrm{th}}$ Decoder layer};
\node (sf3) [emb]    at (3.5, 4.0) {$\mathrm{softmax}$};
\node (pd3) [var]    at (3.5, 5.1) {$\mathcal{L}_{D}^{\text{Attn.}}$};

\node (ca)  [var]    at (1.2, 2.0) {C/A};
\node (sn) [hidden]  at (-0.1, 1.7)   {};
\node (a0)  [hidden] at (1.7, 0.5) {};
\node (a1)  [hidden] at (1.7, 1.7) {};
\node (a2)  [hidden] at (1.7, 3.0) {};

\node (i0)  [hidden] at (5.3, 2.3) {};
\draw [arr] (x)   edge (enc);
\draw [arr] (enc) edge (ctc);
\draw [arr] (ctc) edge (pctc);
\draw [arr] (sf1) edge (pd1);
\draw [arr] (sf3) edge (pd3);

\draw (sn) edge (a1);
\draw (a0) edge (a1);
\draw (a1) edge (a2);
\draw (h13) edge (i0);
\draw [arr] (a0) edge (dec0);
\draw [arr] (dec0) edge[dotted] (dec1);
\draw [arr] (dec1) edge[dotted] (dec2);
\draw [arr] (a1) edge (dec1);
\draw [arr] (a2) edge (dec2);
\draw [arr] (dec2) edge (sf3);

\draw [arr] (i0) edge (sf1);
\end{tikzpicture}
\caption{Architecture of the proposed DeCRED.
In addition to the standard encoder-decoder framework for ASR ($\mathcal{L}_D^{\text{Attn}}$), with the auxiliary CTC objective ($\mathcal{L}^\text{CTC}$), DeCRED uses -- possibly multiple -- auxiliary classifiers ($\mathcal{L}_d^{\text{Attn}}$) attached to the decoder. In the illustration, we show one auxiliary classifier attached to the $(D\text{-}2)$-th decoder block. The embedding and positional encoding layers are not depicted for brevity.}
\label{fig:framework_architecture}
\end{figure}
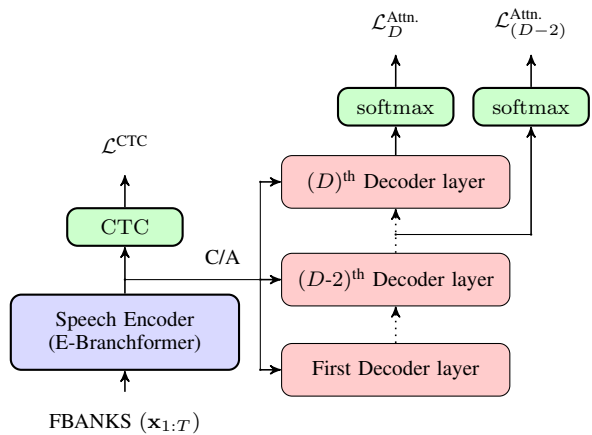




In this work, we present Decoder-Centric Regularisation in Encoder-Decoder (DeCRED), which integrates auxiliary classifiers into intermediate decoder layers and trains them with the same ASR prediction objective as the final layer.
This enforces additional supervision on the decoder with negligible computational overhead (only the extra cost of a single linear layer and the loss function evaluation) during training.
Furthermore, it incurs no additional cost during inference as only the final decoder layer is used for inference (although we note that further improvements in ASR can be obtained by fusing the probabilities from various decoder layers).
Thus, our approach differs from and complements prior approaches in the following ways:
\begin{itemize}
    \item We exclusively regularize the decoder by introducing auxiliary classifiers into intermediate decoder layers.
    \item We analyze the impact of the proposed regularization scheme, showing that it improves the decoder's capacity as an ILM, and that this improvement extends to domains which were not seen during training. Moreover, the ILM gains are not simply theoretical, but also accompanied by both in-domain and out-of-domain ASR improvements.
\end{itemize}

We evaluate DeCRED in a relatively large-scale setup using the E-Branchformer architecture~\cite{kim_e-branchformer_2023}, training on a mix of datasets while reserving some for out-of-domain evaluation.
Our experiments show that DeCRED improves out-of-domain performance on four unseen datasets (AMI~\cite{AMI}, FLEURS~\cite{conneau_fleurs_2023}, Gigaspeech~\cite{gigaspeech} and Earnings-22~\cite{delrio2022earnings22practicalbenchmarkaccents}), reducing the macro-average WER from 18.2\,\% to 16.2\,\%, while also lowering in-domain macro-average WER from 6.4\,\% to 6.3\,\% compared to the non-regularized baseline.
Furthermore, DeCRED complements large-scale training strategies, performing on par with OWSM v3.1 and achieving WERs near Whisper-medium, despite relying on a fraction of their training data and model size.

\section{Decoder-centric regularization}
\label{sec:decred}
Our approach extends the training objective of the hybrid CTC-attention-based training scheme~\cite{hori_joint_2017} by adding auxiliary cross-entropy loss functions. The objective function $\mathcal{L}$ is defined as:
\begin{equation}
\mathcal{L} = \alpha \,\mathcal{L}^{\text{CTC}} + (1 - \alpha)\, \mathcal{L}^{\text{DeCRED}},
\end{equation}
where $\mathcal{L}^{\text{CTC}}$ represents the standard CTC loss~\cite{graves_connectionist_2006}, $\alpha$ is a hyper-parameter, and 
\begin{equation}
    \mathcal{L}^{\text{DeCRED}} = \sum_{d=1}^{D} \beta_{d} \mathcal{L}^{\text{Attn}}_{d},
\end{equation}
where $D$ represents the number of layers in the decoder, $\mathcal{L}^{\text{Attn}}_{d}$ is the cross-entropy loss given a classifier layer (linear projection, followed by softmax function) attached to the $d$-th layer of the decoder, and $\beta_{d}$ is the weighting factor of $d$-th layer. We impose constraints $\sum_{d=1}^{D} \beta_{d} = 1$ and $\beta_{d} \geq 0$. In practise $[\beta_1 \ldots \beta_D]$ is a sparse vector.

This definition allows us to explicitly regularize the ILM and force earlier layers of the decoder to learn discriminative features suitable for the task.
Figure~\ref{fig:framework_architecture} illustrates the proposed architecture, where an auxiliary classifier is attached to the output of $(D\text{-}2)$-th decoder block.

\label{sec:method_decoding}
The decoding follows a typical auto-regressive scheme observed in encoder-decoder ASR systems, where the posterior probability of an output token is obtained by conditioning on previously decoded tokens (partial hypothesis) and the input features.
Formally, let $\mathbf{x}_{1:T}$ be a sequence of input speech (filterbank) features and 
let $y_{1:N}$ be a sequence of output tokens.
Following the joint CTC/attention decoding~\cite{hori_joint_2017}, the log posterior probability of output token $y_n$ is evaluated as
\begin{align}
 \label{eq:decoding}
 {}& \log p(y_n \mid y_{1:n-1}, \mathbf{x}_{1:T}) \approx \lambda \, \log p_{\text{CTC}}(y_n\mid y_{1:n-1},\mathbf{x}_{1:T}) \, + \, \nonumber \\
 &\,\, (1 -\lambda) \, \log p_{\text{DeCRED}}(y_n\mid y_{1:n-1},\mathbf{x}_{1:T}), 
 \end{align}
 where $\lambda$ is a hyper-parameter. 
Now, let $\mathbf{h}_{d,n} \in \mathbb{R}^{1 \times d_{\text{model}}}$ denote the hidden representation corresponding to the $n$-th output token obtained from the $d$-th layer of the decoder, and $\mathbf{W}_{d} \in \mathbb{R}^{d_{\text{model}} \times V}$ represent linear projection from hidden dimension $d_{\text{model}}$ to vocabulary size $V$. 
We obtain the following decoding methods by varying the definition of $p_{\text{DeCRED}}$:

\begingroup
\begin{enumerate}
    \item Vanilla joint CTC/attention decoding relying on representations \textit{only} from the last layer $D$, using auxiliary classifier(s) only as regularization during training:
    \begin{equation}\label{eq:decoding-baseline}
         p_{\text{DeCRED}}(y_n\mid y_{1:n-1},\mathbf{x}_{1:T}) = 
         \text{softmax}(\mathbf{h}_{D,n} \mathbf{W}_{D})
    \end{equation}
    \item  Sum of logits weighted by  per-layer learnable vector $\mathbf{v}_{d} \in \mathbb{R}^{1 \times V}$, where $\odot$ is elementwise product:
    \begin{flalign}\label{eq:decoding-per-token}
         p_{\text{DeCRED}}(\cdot) =  \text{softmax}\left (\sum_{d=1}^{D} \mathbf{v}_{d} \odot (\mathbf{h}_{d,n} \mathbf{W}_{d}) \right)
    \end{flalign}
\end{enumerate}
\endgroup
These schemes can be easily integrated into any decoding search algorithm, such as greedy or beam search.

\begin{table*}[ht]
    \centering
    \caption{
    WERs (with confidence intervals) of ED and DeCRED models reported across multiple in-domain test sets using normalized transcripts, with Whisper-medium and OWSM v3.1 included as references. 
    }
    \label{tab:ed_decred}
    \begin{NiceTabular}{@{}lcccccccc@{}}
        \toprule
        \diagbox{{Model}}{{Dataset}} & {CV-13} & {SB eval2000} & {LS clean} & {LS other} & {TEDLIUM3} & {VoxPopuli} & {WSJ} & {Macro Avg.} \\ 
        \midrule
                

        

        $\text{ED}^{(\ref{eq:decoding-baseline})}$ &
        $\mathbf{11.9}~_{11.7}^{12.2}$ & 
        $9.2~_{8.8}^{9.6}$ & 
        $2.5~_{2.3}^{2.7}$ & 
        $5.7~_{5.4}^{6.1}$ & 
        $6.6~_{6.1}^{7.1}$ & 
        $7.5~_{6.9}^{8.2}$ & 
        $1.8~_{1.5}^{2.2}$ & 
        $6.4$ \\

        $\text{DeCRED}^{(\ref{eq:decoding-baseline})}$ & 
        $12.0~_{11.8}^{12.3}$ & 
        $9.4~_{8.9}^{9.7}$ & 
        $2.4~_{2.2}^{2.5}$ & 
        ${5.5}~_{5.3}^{5.7}$ & 
        $6.3~_{5.9}^{6.8}$ & 
        $\mathbf{7.3}~_{6.8}^{8.0}$ & 
        $\mathbf{1.5}~_{1.2}^{1.9}$ & 
        $6.3$ \\

        $\text{DeCRED}^{(\ref{eq:decoding-per-token})}$ &
        $12.2~_{11.8}^{12.5}$ & 
        $\mathbf{9.1}~_{8.7}^{9.5}$ & 
        $\mathbf{2.3}~_{2.2}^{2.5}$ & 
        ${5.5}~_{5.3}^{5.8}$ & 
        $5.7~_{5.3}^{6.1}$ & 
        $\mathbf{7.3}~_{6.8}^{7.9}$ & 
        $\mathbf{1.5}~_{1.2}^{1.8}$ & 
        $\mathbf{6.2}$ \\

        \midrule
        Whisper medium &
        $12.4~_{12.1}^{12.6}$ & 
        $14.7~_{14.2}^{15.2}$ & 
        $3.0~_{2.7}^{3.4}$ & 
        $5.9~_{5.6}^{6.2}$ & 
        $\mathbf{4.2}~_{3.8}^{4.6}$ & 
        $8.0~_{7.4}^{8.8}$ & 
        $3.2~_{2.6}^{3.8}$ & 
        $7.3$ \\

        OWSM v3.1 & 
        $12.9~_{12.5}^{13.2}$ & 
        $11.2~_{9.2}^{14.6}$ & 
        $2.4~_{2.2}^{2.6}$ & 
        $\mathbf{5.0}~_{4.8}^{5.3}$ & 
        $5.0~_{4.7}^{5.4}$ & 
        $8.5~_{8.0}^{9.0}$ & 
        $3.5~_{2.9}^{4.0}$ & 
        $6.9$ \\

                



        
        \bottomrule
    \end{NiceTabular}
\end{table*}

\section{Experimental setup}
\label{sec:exp}

\subsection{Model architecture}
Our baseline Encoder-Decoder (ED) model consists of 16 E-Branchformer~\cite{kim_e-branchformer_2023} encoder layers with relative positional embeddings~\cite{dai_transformer-xl_2019}, Macaron-like feedforward modules~\cite{gulati_conformer_2020}, $d_{\text{model}}= 512$,  $d_{\text{ff}} = 4d_{\text{model}}$, four attention heads, and a dropout probability of 0.1.
In line with the E-Branchformer architecture, we incorporate a merge block followed by depth-wise convolution with a kernel size of 31. The encoder is followed by an 8-layer Transformer decoder with sinusoidal positional embeddings, maintaining the same number of attention heads, $d_{\text{model}}$, $d_{\text{ff}}$, and dropout ratio. 

The ED model has 172M parameters and processes 80-dimensional filter-bank features as input. These first pass through two 2D convolutional layers with 512 output channels, a kernel size of $3\times3$, and a stride of $2\times2$, reducing sequence length. A linear projection then matches $d_{\text{model}}$. We use a subword tokenizer with a vocabulary of size $V=5000$ based on the Unigram algorithm from~\cite{kudo_sentencepiece_2018}.
Unless stated otherwise, the DeCRED model extends the baseline ED model by adding a single auxiliary classifier with $\beta_{D-2} = 0.4$, introducing just $d_{\text{model}} \times V$ additional parameters. By default joint greedy decoding with ($\lambda = 0.3$) is utilized.
\subsection{Datasets}
Following the data selection strategy of SpeechStew~\cite{chan_speechstew_2021}, we construct our training set using a mixture of diverse, multi-domain speech datasets: Fisher (SWITCHBOARD)~\cite{godfrey_switchboard_1992},  WSJ~\cite{paul_design_1992}, Common Voice en 13~\cite{ardila_common_2020}, LibriSpeech~\cite{panayotov_librispeech_2015}, VoxPopuli~\cite{wang_voxpopuli_2021}, and TED-LIUM~3~\cite{hernandez_ted-lium_2018}, amounting to roughly 6,000 hours of transcribed audio.
As reported in~\cite{peng_reproducing_2023}, training on such heterogeneous datasets risks overfitting to domain-specific annotation styles, which may hinder general-purpose performance. To mitigate this, we standardize all transcripts using the Whisper normalizer,\footnote{\url{https://github.com/openai/whisper/blob/main/whisper/normalizers/english.py}}, ensuring consistency across datasets and reducing annotation-style biases.

To evaluate the generalization ability of our models, we test them on four datasets not seen during training: AMI~\cite{AMI}, FLEURS~\cite{conneau_fleurs_2023}, GigaSpeech~\cite{gigaspeech}, and Earnings-22~\cite{delrio2022earnings22practicalbenchmarkaccents}.

\subsection{Training setup}
All experiments are conducted using the open-source \texttt{transformers} library and trained on Nvidia A100 GPUs with the AdamW optimizer~\cite{loshchilov2018decoupled}. Training runs for 100 epochs with early stopping (patience of 10), a learning rate of $2\times 10^{-3}$, weight decay of $1\times 10^{-6}$, a linear decay scheduler, 40k warm-up steps, and label smoothing~\cite{PereyraTCKH17, watanabe_espnet_2018} with a weight of 0.1. To accelerate training, samples longer than 20 seconds are discarded.

We apply speed perturbation with randomly selected factors ${0.9, 1.0, 1.1}$ and delay SpecAug~\cite{park_specaugment_2019} until after 5k update steps. For all experiments, the best-performing checkpoint is selected based on the development WER.

Furthermore, we introduce a mechanism to mask special tokens and unfinished words (e.g., transcript ``[hesitation] to re- to re- renew" is transformed into ``[MASK] to [MASK] to [MASK] renew") during error backpropagation by not reflecting [MASK] token in the loss function. This strategy prevents penalization for unclear inputs, which are particularly common in the Fisher dataset.


\section{Results}
\label{sec:multi-domain}
We evaluate our models using WER with confidence intervals\footnote{Confidence intervals are displayed as subscripts and superscripts in corresponding tables.} computed via bootstrapping ($\alpha = 0.05$, $B = 1000$)~\cite{Confidence_Intervals}. Significance tests are performed with pair-wise bootstrapping. 

\subsection{In domain performance}

Table~\ref{tab:ed_decred} presents the WER results of the baseline $\text{ED}^{(\ref{eq:decoding-baseline})}$ and the proposed $\text{DeCRED}^{(\ref{eq:decoding-baseline})}$ models across in-domain datasets. Superscripts in model names indicate the decoding strategy used—specifically, Equation~\eqref{eq:decoding-baseline} corresponds to decoding from the last layer of the model.

Notably, $\text{DeCRED}^{(\ref{eq:decoding-baseline})}$ outperforms $\text{ED}^{(\ref{eq:decoding-baseline})}$ on 5 out of 7 datasets. 
$\text{ED}^{(\ref{eq:decoding-baseline})}$ achieves lower WER on CV-13, with a $p$-value of 0.26 when compared to $\text{DeCRED}^{(\ref{eq:decoding-baseline})}$. Similarly, on Switchboard (eval2000), it outperforms $\text{DeCRED}^{(\ref{eq:decoding-baseline})}$ with a $p$-value of 0.19. On all other test sets, $\text{DeCRED}^{(\ref{eq:decoding-baseline})}$ demonstrates improvements over the baseline, with the following $p$-values: LS clean (0.24), LS other (0.13), TED-LIUM3 (0.20), VoxPopuli (0.19), and WSJ (0.10).

Furthermore, inspired by Platt scaling~\cite{Guo_calibrating_2017, Lee_calibration_2021}, we freeze the model parameters and train only the mixing weights $\mathbf{v}$ for the decoding method described in Equation~\eqref{eq:decoding-per-token}. With this setup, $\text{DeCRED}^{(\ref{eq:decoding-per-token})}$ outperforms $\text{ED}^{(\ref{eq:decoding-baseline})}$ on 6 out of 7 datasets (except CV-13), with a $p$-value of 0.4 for Switchboard eval2000 and $p$-values below 0.2 for the remaining datasets.


\begingroup
\setlength{\tabcolsep}{2pt} 
\begin{table}[t]
    \centering
    \caption{%
        Comparison of ED and DeCRED models on out-of-domain test sets.
    }\label{tab:ood}
    \small{    \begin{NiceTabular}{
        m{2cm} 
        >{\centering\arraybackslash}m{1.15cm} 
        >{\centering\arraybackslash}m{1.15cm} 
        >{\centering\arraybackslash}m{1.15cm} 
        >{\centering\arraybackslash}m{1.15cm} 
        >{\centering\arraybackslash}m{1.15cm}
    }
        \toprule
        \diagbox{{Model}}{{Dataset}}& FLEURS & AMI ihm& Giga-speech & Earnings-22 & Macro Avg. \\
        \midrule
        $\text{ED}^{(\ref{eq:decoding-baseline})}$ & 6.4$~_{5.9}^{6.9}$ & 24.8$~_{23.4}^{26.5}$ & 20.1$~_{19.6}^{20.7}$ & 21.4$~_{19.7}^{23.5}$ & 18.2 \\
        $\text{DeCRED}^{(\ref{eq:decoding-baseline})}$ & 6.7$~_{6.2}^{7.1}$ & 22.1$~_{21.6}^{22.7}$ & 16.9$~_{16.6}^{17.3}$ & 19.0$~_{18.3}^{19.8}$ & 16.2 \\
        $\text{DeCRED}^{(\ref{eq:decoding-per-token})}$ & 6.7$~_{6.2}^{7.3}$ & 21.9$~_{21.5}^{22.3}$ & 16.7$~_{16.4}^{17.0}$ & 18.3$~_{17.6}^{19.0}$ & 15.9 \\
        \midrule
        OWSM v3.1 & 7.2$~_{6.7}^{7.8}$ & 23.3$~_{19.8}^{26.9}$ & 19.2$~_{17.9}^{20.4}$ & 14.0$~_{13.5}^{14.5}$ & 15.9 \\
        Whisper medium & \textbf{4.5}$~_{4.2}^{4.9}$ & \textbf{16.6}$~_{16.0}^{17.6}$ & \textbf{13.8}$~_{12.6}^{15.4}$ & \textbf{11.7}$~_{11.2}^{12.2}$ & 11.7 \\
        \bottomrule
    \end{NiceTabular}}
\end{table}
\endgroup

\begingroup
\setlength{\tabcolsep}{2.5pt} 
\begin{table*}[ht]
    \centering
    \caption{%
        Zero-Attention ILM BPE-level perplexity estimation of ED and DeCRED models on in- and out-of-domain test sets.
    }\label{tab:ppl}
    \small{\begin{NiceTabular}{lccccccc|cccc}
        \toprule
        \diagbox{{Model}}{{Dataset}}  & CV-13 & LS clean & LS other & SB eval2000 & TEDLIUM3 & VoxPopuli & WSJ & FLEURS & AMI-ihm & Gigaspeech & Earnings-22  \\
        \midrule
        $\text{ED}^{(\ref{eq:decoding-baseline})}$    & 455.8 & 459.8 & 473.3 & 474.0 & 297.6 & 286.2 & 676.8 & 306.7 & 537.8 & 297.7 & 592.1\\         
        $\text{DeCRED}^{(\ref{eq:decoding-baseline})}$ & 215.7 & 209.0 & 197.5 & 271.6 & 140.4 & 141.0 & 723.2 & 161.1 & 310.4 & 134.1 & 266.7 \\
        \bottomrule
    \end{NiceTabular}}
\end{table*}
\endgroup

To provide better context for the reader, we include a reference comparison with large-scale multilingual models: Whisper-medium\footnote{We use the multilingual version as it performs better across our datasets.}~\cite{radford_robust_2023} (764M parameters) and OWSM v3.1~\cite{peng24b_interspeech} (1.02B parameters). This is intended as a point of reference rather than a direct comparison, given the substantial differences in model scale and design. For consistency, we apply the same text normalization pipeline to the outputs of Whisper and OWSM in our normalized setup. We also trained smaller variants of ED-small and DeCRED-small with 39M parameters, achieving macro-average WERs of 8.4\,\% and 8.1\,\%, respectively.

\subsection{Out of domain performance}
Table~\ref{tab:ood} underscores the significant WER reductions achieved by $\text{DeCRED}^{(\ref{eq:decoding-baseline})}$ and $\text{DeCRED}^{(\ref{eq:decoding-per-token})}$ on out-of-domain datasets, highlighting a major outcome of our work. Despite not being exposed to these datasets during training, DeCRED delivers a macro WER reduction of 2.0 and 2.3 percentage points, respectively.

While $\text{ED}^{(\ref{eq:decoding-baseline})}$ performs better on FLEURS ($p$-values: 0.13 for $\text{DeCRED}^{(\ref{eq:decoding-baseline})}$ and 0.24 for $\text{DeCRED}^{(\ref{eq:decoding-per-token})}$). $\text{DeCRED}^{(\ref{eq:decoding-baseline})}$ and $\text{DeCRED}^{(\ref{eq:decoding-per-token})}$ achieve statistically significantly lower WER on AMI ($p = 0.004$, $<0.001$), Gigaspeech ($p < 0.001$, $<0.001$), and Earnings-22 ($p = 0.04$, 0.13)  respectively.

Notably, OWSM v3.1 was trained on FLEURS, AMI, and Gigaspeech, while Whisper was trained on web-scale data that may include these datasets. Despite this, DeCRED performs comparably to OWSM v3.1 and remains competitive with Whisper, even when using greedy decoding with $\lambda=0$.


\subsection{Analysis of internal language model}
In Table~\ref{tab:ppl}, we present a comparison of the subword-level perplexity estimates of the Zero-Attention Internal Language Model (ILM)~\cite{zeineldeen_investigating_2021} for the ED and DeCRED models. We note that these estimates should be interpreted with caution, as the ILM estimate is approximate and is not guaranteed to be valid when the end-to-end model does not strictly satisfy the conditions outlined in Proposition 1 of~\cite[Appendix A]{HAT},~\cite{ILME}. With this caveat, the table shows consistent reductions in ILM perplexity for DeCRED compared to ED across all evaluated datasets. These reductions suggest an improvement in the regularization or internal language modeling capacity of DeCRED, which aligns with the WER trends observed in Table~\ref{tab:ed_decred} and Table~\ref{tab:ood}.



\begin{figure}[t]
    \centering
    \includegraphics[width=\linewidth]{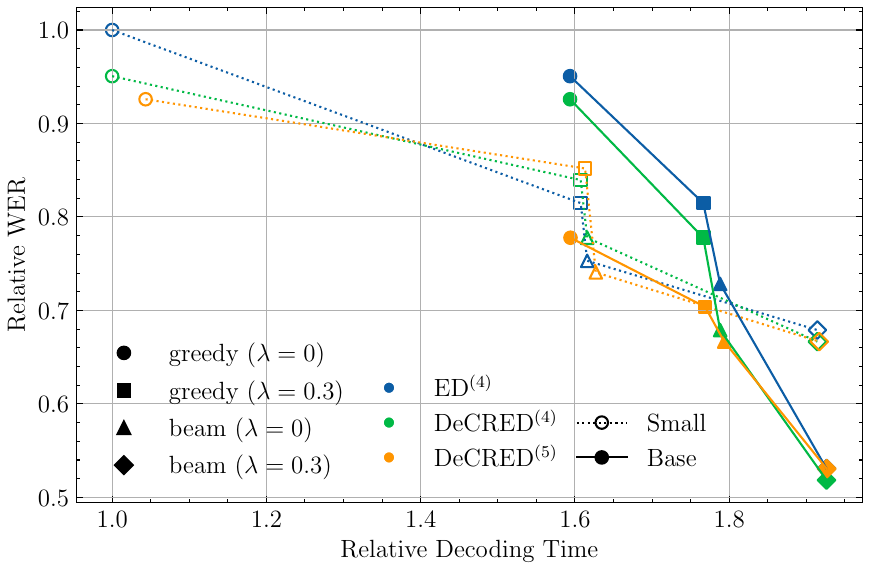}
    \caption{%
        The impact of model size and decoding approach on the average time needed to transcribe an utterance and macro average WER}\label{fig:slowdown_wer}
\end{figure}

\begingroup
\setlength{\tabcolsep}{3pt} 
\begin{table}[t]
    \centering    
    \caption{%
    Macro average WERs on in-domain datasets for different decoding strategies.
    }\label{tab:decoding_analysis}
    \small{
    \begin{NiceTabular}{cccc}
    \toprule
    
    Decoding                      & \multicolumn{2}{c}{greedy} & \multicolumn{1}{c}{beam}       \\
    strategy                          & $\lambda=0$   & $\lambda=0.3$   & width 10, $\lambda=0.3$    \\
    \midrule
    $\text{DeCRED}^{(\ref{eq:decoding-baseline})}$                              & 6.7   & 6.3   & \textbf{5.8} \\
    Early exiting                             & 6.7    & 6.6    & 6.2\\
    $\text{DeCRED}^{(\ref{eq:decoding-per-token})}$                            & \textbf{6.5}   & \textbf{6.2}   & \textbf{5.8}             \\
    \bottomrule
    \end{NiceTabular}
    }
\end{table}
\endgroup

\section{Further analysis}
\subsection{Trade-off between performance and decoding time}
Table~\ref{tab:decoding_analysis} shows macro-average WERs on in-domain datasets for various decoding strategies applied to DeCRED. Early exiting~\cite{scardapane_why_2020} is a special case of $\text{DeCRED}^{(\ref{eq:decoding-per-token})}$, where $\mathbf{v}_{D-2}[:]$ is set to one, and $\mathbf{v}_D[:]$ are zeros. This allows decoding from a fixed intermediate layer to reduce computation cost.

The improvements in WER of $\text{DeCRED}^{(\ref{eq:decoding-per-token})}$ are most notable under greedy decoding, which is illustrated in Figure~\ref{fig:slowdown_wer}. The relative slowdown is measured against the fastest model, $\text{ED-small}^{(\ref{eq:decoding-baseline})}$, using a fixed number of decoding steps and batch sizes constrained by the GPU memory of the A100 GPU.

As shown in the figure, DeCRED improves WER over ED with minimal overhead. For $\text{DeCRED}^{(\ref{eq:decoding-per-token})}$, the additional cost comes only from computing $\mathrm{softmax}\left (\sum_{d=1}^{D} \mathbf{v}_{d} \odot (\mathbf{h}_{d} \mathbf{W}_{d}) \right)$. In particular, DeCRED-small  (\textcolor{C2}{$\circ$}) with greedy decoding performs similarly to ED (\textcolor{C0}{\textbullet}), while being more efficient and requiring fewer resources.

\begin{table*}[ht]
    \centering
    \caption{
        WERs (together with confidence intervals) of $\text{DeCRED}^{(\ref{eq:decoding-baseline})}$, Whisper medium, and OWSM v3.1 models evaluated on original transcripts across multiple in-domain test sets.
    }
    \label{tab:normalization_effect}
    \begin{NiceTabular}{@{}lcccccccc@{}}
        \toprule
        \diagbox{{Model}}{{Dataset}} & {CV-13} & {SB eval2000} & {LS clean} & {LS other} & {TEDLIUM3} & {VoxPopuli} & {WSJ} & {Macro Avg.} \\ 
        \midrule

        Whisper medium & 
        $\mathbf{13.2}~_{12.6}^{13.9}$ & 
        $28.9~_{27.5}^{30.6}$ & 
        $3.9~_{3.6}^{4.2}$ & 
        $7.0~_{6.7}^{7.5}$ & 
        $6.0~_{5.6}^{6.4}$ & 
        $10.5~_{10.0}^{11.2}$ & 
        $9.4~_{8.0}^{10.7}$ & 
        $11.3$ \\

        $\text{DeCRED}^{(\ref{eq:decoding-baseline})}$ & 
        $15.0~_{14.4}^{15.9}$ & 
        ${22.7}~_{21.7}^{23.8}$ & 
        $3.8~_{3.5}^{4.1}$ & 
        $7.3~_{6.9}^{7.7}$ & 
        $\mathbf{5.6}~_{5.1}^{6.0}$ & 
        $\mathbf{8.4}~_{7.8}^{9.2}$ & 
        $\mathbf{3.0}~_{2.6}^{3.3}$ & 
        ${9.4}$ \\
    
        OWSM v3.1 & 
        $14.3~_{14.0}^{14.6}$ & 
        $\mathbf{22.3}~_{20.3}^{25.8}$ & 
        $\mathbf{2.6}~_{2.4}^{2.8}$ & 
        $\mathbf{5.3}~_{5.1}^{5.5}$ & 
        $6.1~_{5.7}^{6.5}$ & 
        $9.6~_{9.1}^{10.2}$ & 
        $4.7~_{4.1}^{5.3}$ & 
        $\mathbf{9.3}$ \\
        
        \bottomrule
    \end{NiceTabular}
\end{table*}

\subsection{Effect of text normalization}
For consistency, we report most of the results in this work using normalized transcripts. However, to allow a fair and direct comparison with previous and future works, we additionally trained and evaluated $\text{DeCRED}^{(\ref{eq:decoding-baseline})}$ on original (unnormalized) transcripts. Table~\ref{tab:normalization_effect} presents the resulting WERs, comparing DeCRED with Whisper medium and OWSM v3.1.

DeCRED achieves performance on par with OWSM v3.1, with a comparable macro-average WER (9.4 vs. 9.3), despite operating at a significantly smaller scale—172M parameters vs. 1.5B, 6K hours of English vs. 180K hours of multilingual data, and 2.2K vs. 24.6K A100 GPU hours. These results underscore the effectiveness of the proposed decoder-side regularization method, accompanied by an efficient training pipeline. We publicly release our recipes and framework to support reproducibility and adoption\footnote{\url{https://github.com/BUTSpeechFIT/DeCRED}}. 
For a broader comparison with more state-of-the-art models, we refer the reader to \emph{Open Automatic Speech Recognition Leaderboard}~\cite{open-asr-leaderboard}.

\begin{table}[t]
\centering
\caption{
WER comparison of our ED implementation vs. ESPnet's ED, and DeCRED vs. InterCTC on the TEDLIUM3 test set.
}\label{tab:espnet_v_hf}
\small{
\begin{NiceTabular}{lccc}
\toprule
\text{Model} & Size [M] & greedy & beam -- width 40 \\
\midrule
ESPnet $\text{ED}^{(\ref{eq:decoding-baseline})}$ & 35.01 & 8.7 & 8.1 \\
\midrule
Our $\text{ED}^{(\ref{eq:decoding-baseline})}$ & 35.04 & 7.6 &7.2 \\
$\text{InterCTC}^{(\ref{eq:decoding-baseline})}_{\lfloor L/2 \rfloor}$ & 35.20 & 7.5 & 7.1 \\
$\text{DeCRED}^{(\ref{eq:decoding-baseline})}$ & 35.20 & \textbf{7.0} & \textbf{6.8} \\
\bottomrule
\end{NiceTabular}
}
\end{table}

\begingroup
\setlength{\tabcolsep}{5pt} 
\begin{table}[t]
    \centering
    \caption{%
        Effect of the auxiliary classifier's position ($d$) and weight ($\beta_d$) on WER for the TEDLIUM3 test set.
    }\label{tab:location_weight}
    \small{
    \begin{NiceTabular}{cllll}
    \toprule
    \multirow{2}{*}{\text{Weight} $\beta_d$} & \multicolumn{4}{c}{\text{Position} $d$}\\
    \cmidrule(l){2-5}
     & 2     & 3     & 4     & 5     \\
    \midrule
    0.3 &       & 7.0 & 7.0  & 7.0 \\
    0.4 & 7.5  & 7.1  & 6.8 & 6.9 \\
    0.5 & 7.1  & \textbf{6.7} & 7.1  & 6.9 \\
    \bottomrule
    \end{NiceTabular}
    }
\end{table}
\endgroup

\subsection{Comparison with Encoder-Centric Regularization}
To contextualize DeCRED among related encoder-decoder regularization approaches, we conduct additional experiments on the TEDLIUM3 dataset using small models (37M parameters) for both ED and DeCRED.

Table~\ref{tab:espnet_v_hf} compares DeCRED with the ED baseline and \textit{InterCTC}~\cite{lee_intermediate_2021}, a related method that applies intermediate supervision to the encoder. Both InterCTC and DeCRED introduce a single auxiliary classifier with exactly the same parameter overhead ($d_{\text{model}} \times V$), but only differ in the point of application: InterCTC regularizes the encoder, while DeCRED targets the decoder module. Although both approaches outperform the baseline ED, DeCRED yields the lowest WER, particularly under greedy decoding.

For completeness, we also report results for the ESPnet ED baseline\footnote{\url{https://github.com/espnet/espnet/tree/master/egs2/tedlium3/asr1}}, whose architecture, training configuration, and decoding setup we closely replicated to ensure a comparable evaluation.

Finally, Table~\ref{tab:location_weight} examines the effect of varying the auxiliary classifier’s position ($d$) and loss weight ($\beta_d$). Consistent improvements are observed when placing the classifier around decoder layers 3 or 4 (the latter corresponding to $D{-}2$), aligning with InterCTC’s findings on optimal supervision depth. The best performance is achieved with $\beta_3 = 0.5$ (6.7\%) and $\beta_4 = 0.4$ (6.8\%), indicating that even a single well-placed decoder-side classifier is sufficient to obtain strong gains. Adding multiple auxiliary classifiers did not lead to any significant improvements.

\section{Conclusion and limitations}
We introduced the DeCRED regularization scheme, which effectively integrates an auxiliary classifier within the decoder of an encoder-decoder architecture. Alongside this, we proposed a novel decoding method that leverages these classifiers. Without any additional computational overhead, DeCRED achieved lower WERs than the baseline model on 5 out of 7 in-domain datasets. More importantly, on 3 out of 4 out-of-domain datasets, DeCRED obtained statistically significant WER reductions compared to the baseline. On average, DeCRED reduced the WER by 2.0 absolute percentage points on out-of-domain datasets, and the proposed domain adaptation scheme further improved WER by 0.3 absolute points in this setting. Despite its relatively small size, DeCRED achieved WERs comparable to much larger models such as OWSM v3.1 and Whisper medium.

Despite these promising results, we identify a few limitations in our work. First, due to computational budget constraints, we were only able to scale our experiments to 6k hours of training data and to a model with 172M parameters. Secondly, our models were trained exclusively on English data, which complicates direct comparisons with multilingual models, as these models must allocate part of their capacity to handle multiple languages. 
It is also worth noting that some of the improvements from DeCRED diminish when beam-search decoding with a wide beam is employed, as this comes at a computational cost during inference. 


\section*{Acknowledgements}
The work was supported by Ministry of Education, Youth and Sports of the Czech Republic (MoE) through the OP JAK project ``Linguistics, Artificial Intelligence and Language and Speech Technologies: from Research to Applications" (ID:CZ.02.01.01/00/23\_020/0008518). Computing on IT4I supercomputer was supported by MoE through the e-INFRA CZ (ID:90254).

\bibliographystyle{IEEEtran}
\bibliography{references}

\end{document}